%% file: main.tex
\documentclass[twocolumn,showpacs,superscriptaddress,email,floatfix,longbibliography,prl]{revtex4-2}
\usepackage{graphicx}
\usepackage{dcolumn}
\usepackage{bm}
\usepackage{amsmath}
\usepackage{bbold}
\usepackage{xcolor}
\usepackage[hypertexnames=false]{hyperref}
\usepackage{physics} 

\usepackage{orcidlink}
\usepackage{subfiles} 

\begin{document}

\preprint{APS/123-QED}

\title{Dispersionless subradiant photon storage in one-dimensional emitter chains}

\author{Marcel Cech \,\orcidlink{0000-0002-9381-6927}}
\affiliation{Institut f\"ur Theoretische Physik, Universit\"at Tübingen, Auf der Morgenstelle 14, 72076 T\"ubingen, Germany}
\author{Igor Lesanovsky \,\orcidlink{0000-0001-9660-9467}}
\affiliation{Institut f\"ur Theoretische Physik, Universit\"at Tübingen, Auf der Morgenstelle 14, 72076 T\"ubingen, Germany}
\affiliation{School of Physics and Astronomy and Centre for the Mathematics and Theoretical Physics of Quantum Non-Equilibrium Systems, The University of Nottingham, Nottingham, NG7 2RD, United Kingdom}
\author{Beatriz Olmos \,\orcidlink{0000-0002-1140-2641}}
\affiliation{Institut f\"ur Theoretische Physik, Universit\"at Tübingen, Auf der Morgenstelle 14, 72076 T\"ubingen, Germany}
\affiliation{School of Physics and Astronomy and Centre for the Mathematics and Theoretical Physics of Quantum Non-Equilibrium Systems, The University of Nottingham, Nottingham, NG7 2RD, United Kingdom}

\begin{abstract}
Atomic emitter ensembles couple collectively to the radiation field. Although an excitation on a single emitter may be short-lived, a collection of them can contain a photon several orders of magnitude longer than the single emitter lifetime. We provide the exact conditions for optimal absorption, long-lived and dispersionless storage, and release, of a single photon in a sub-wavelength one-dimensional lattice of two-level emitters. In particular, we detail two storage schemes. The first is based on the uncovering of approximate flat sections in the single-photon spectrum, such that a single photon can be stored as a wave packet with effective zero group velocity. For the second scheme we exploit the angular dependence of the interactions induced between the emitters and mediated via exchange of virtual photons, which on a ring gives rise to an effective trapping potential for the photon. In both cases, we are able to obtain, within current experimentally accessible parameters, high-fidelity photon storage for times hundreds of times longer than the single emitter lifetime.
\end{abstract}

\maketitle

\textit{Introduction.} When an ensemble of emitters is coupled to a common environment, its dynamics is subject to collective phenomena \cite{Dicke1954,Lehmberg1970,James1993}. Exchange of virtual photons induces dipole-dipole interactions among the emitters. Moreover, spontaneous emission of photons from the ensemble becomes either enhanced (superradiance) or inhibited (subradiance), depending on the collective state in which the photons are stored in the ensemble. These effects are particularly prominent, dominating the dynamics of the system, when the distance between the emitters is smaller than the wavelength of the considered light \cite{Olmos2013,Pellegrino2014,Jenkins2016,Bromley2016,Manzoni2018,Jenkins2017,MorenoCardoner2019,Zhang2020flatband,Masson2020,Ferioli2021,Ballantine2021,Petrosyan2021,RubiesBigorda2022}, in ensembles with many emitters \cite{Bienaime2012,Ott2013,Araujo2016,Guerin2016}, and in the presence of structured environments, such as nanophotonic waveguides \cite{Ruostekoski2016,Asenjo2017,Lodahl2017,Asenjo2017_1,Ostermann2019,Kornovan2019,Poddubny2022,Pennetta2022,Chu2022}, cavities \cite{Shankar2021,Gegg2018,Pineiro2022,Hotter2023}, and resonators \cite{Kornovan2021}. Owed to the high degree of experimental control available in cold atomic and solid state systems, these collective phenomena are now in the vanguard of research on quantum information processing and metrology \cite{Reitz2022}.

The existence of subradiant, i.e. long-lived, states offers the opportunity of storing light in emitter ensembles for times that exceed the lifetime of a single emitter by several orders of magnitude. Many theoretical and experimental works have been put forward in the last few years aiming to use this phenomenon for photon storage and release \cite{Needham2019,Asenjo2017,Ferioli2021,Ballantine2021,RubiesBigorda2022}, enhanced quantum metrology \cite{Ostermann2013}, atomic optical mirrors \cite{Bettles2016,Rui2020,Ballantine2022,Buckley2022,Srakaew2023}, and entangled state preparation \cite{Guimond2019,Buonaiuto2019}, among others. The main challenge with subradiant states lies in their preparation as, by their very definition, they possess little overlap with typical electric field radiation modes. In a sub-wavelength periodic arrangement of emitters, subradiant states possess intricate phase patterns, making their laser excitation highly involved, requiring, for example, phase imprinting protocols or spatially dependent external electric and magnetic fields \cite{Jen2016,Jen2017,Jen2018,Ballantine2021,RubiesBigorda2022}.

\begin{figure}[t]
    \centering
    \includegraphics{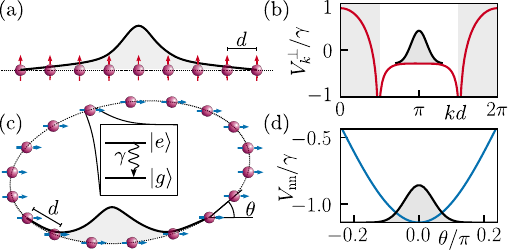}
    \caption{\textbf{Dispersionless photon trapping.} Single photon stored as a subradiant wave packet in a 1D lattice of two-level systems. (a-b): In a chain, the nearest neighbor distance $d$ is chosen such that the wave packet is localised in a region where the dispersion relation is flat and outside the radiative region (gray shaded areas). (c-d): In a ring, the spatial variation of the nearest neighbour interaction $V_\mathrm{nn}$ creates an effective trapping potential for the photon.}
    \label{fig:fig1}
\end{figure}
\begin{figure*}[t]
    \centering
    \includegraphics{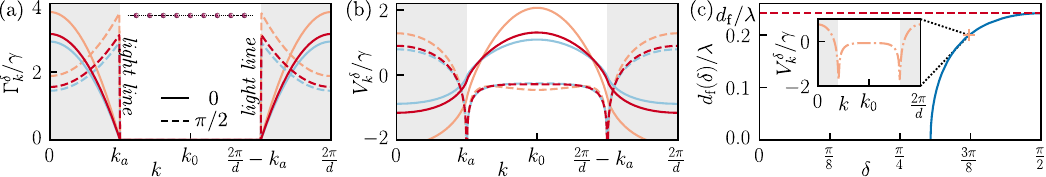}
    \caption{\textbf{Flat dispersion relation.} (a): Decay rates and (b): spectrum for an infinite 1D lattice with $d/\lambda=0.2$ (orange), $0.2414$ (red) and $0.26$ (light blue). Independently of the angle $\delta$ between the dipole moment and the chain, completely subradiant modes ($\Gamma_k^\delta=0$) can be found between the light lines, i.e. for all $2\pi/d-k_a>k>k_a=2\pi/\lambda$. For $\delta=\pi / 2$, the spectrum can become approximately flat around $k=k_0=\pi/d$. (c): Lattice spacing $d_\mathrm{f}(\delta)$ (blue line), where the second derivative of $V_k^\delta$ vanishes at $k = k_0$ giving rise to spectra with a flat region. The maximal lattice spacing with a flat dispersion is $d_\mathrm{f} \equiv d_\mathrm{f}(\pi / 2) = 0.2414 \lambda$ (red dashed line). The inset shows the flat region in the spectrum at $d_\mathrm{f}(3\pi/8) \approx 0.2\lambda$. }
    \label{fig:fig2}
\end{figure*}

In this work, we present two schemes for subradiantly storing a single photon in a one dimensional lattice that overcome the above challenge and that, moreover, allow for the dispersionless storage of the photon. The first is based on the preparation of a subradiant wave packet with zero group velocity by finding an optimal ratio between the lattice constant and the photon wavelength such that a large section of the spectrum displays an approximately flat dispersion relation. The second scheme exploits the strong angular dependence of the dipole-dipole interactions. By placing the emitters on a ring geometry, we find that the spatial variation of the interactions create an effective trapping potential for the photon. By analyzing the eigenstates of this potential, we identify the conditions to ensure the long-lived and high-fidelity photon storage.

\textit{System and master equation.} We consider an ensemble of $N$ emitters trapped in a one dimensional configuration --a chain or a ring-- with lattice constant $d$ (see Fig.~\ref{fig:fig1}). We consider here two-level systems, where the ground and excited states, $\ket{g}$ and $\ket{e}$, respectively, are separated by an energy $\hbar\omega_a=h c/\lambda$. All emitters are coupled to the free radiation field. Under the Born-Markov and secular approximations, the dynamics of the emitters' degrees of freedom, included in the density matrix $\rho$, is described by the master equation
\begin{equation}\label{eq:meq}
    \dot{\rho}=-\frac{\mathrm{i}}{\hbar}\left[H,\rho \right]+\sum_{\alpha, \beta}\Gamma_{\alpha \beta}\left(\sigma_\beta\rho\sigma_\alpha^\dagger-\frac{1}{2}\{\sigma_\alpha^\dagger\sigma_\beta,\rho\}\right) \, ,
\end{equation}
with $H=-\hbar\sum_{\alpha\neq\beta}V_{\alpha \beta}\sigma_\alpha^\dagger \sigma_\beta$,
where $\sigma_\alpha=\ketbra{g_\alpha}{e_\alpha}$ and $\sigma^\dag_\alpha=\ketbra{e_\alpha}{g_\alpha}$. The first term of Eq.~(\ref{eq:meq}) represents the dipole-dipole interactions between emitters which, for emitters $\alpha$ and $\beta$, occur at a rate $V_{\alpha\beta}$. The second term represents the dissipation (spontaneous emission of photons into the radiation field), which possesses a collective character: while the diagonal elements of matrix $\Gamma_{\alpha\beta}$ represent the single-emitter spontaneous decay rate $\Gamma_{\alpha\alpha}=\gamma$, collective single-photon decay modes (the eigenvectors of $\Gamma_{\alpha\beta}$) arise due to the presence of non-zero off-diagonal elements. When the associated decay rate of a collective mode is larger (smaller) than $\gamma$, the mode is said to be superradiant (subradiant).

Given an environment, both dipole-dipole and dissipation rates can be obtained in terms of the real and imaginary part of the Green's tensor $\bar{G}(\mathbf{r}_\alpha,\mathbf{r}_\beta,\omega_a)$ evaluated at the positions $\mathbf{r}_\alpha$ and $\mathbf{r}_\beta$ of the emitters~\cite{Asenjo2017_1,Dzsotjan2010} as $\Gamma_{\alpha \beta} = \frac{2\omega^2_a}{\hbar\epsilon_0c^2}\mathbf{d}^*_\alpha\text{Im}\{\bar{G}(\mathbf{r}_\alpha,\mathbf{r}_\beta,\omega_a)\}\mathbf{d}^T_\beta$ and $V_{\alpha \beta} = \frac{\omega_a^2}{\hbar\epsilon_0c^2}\mathbf{d}^*_\alpha\text{Re}\{\bar{G}(\mathbf{r}_\alpha,\mathbf{r}_\beta,\omega_a)\}\mathbf{d}^T_\beta$, where $\mathbf{d}_\alpha$ is the dipole transition vector corresponding to emitter $\alpha$ (note that we consider all dipoles to be aligned, i.e. $\mathbf{d}_\alpha=\mathbf{d}_\beta\equiv\mathbf{d}$). In this work, we will concern ourselves with the situation where all the emitters are in free-space. Here, the Green's tensor is analytically given by
\begin{eqnarray}\label{eq:G0}
    \bar{G}_0(\mathbf{x}_j,k_a)&=&\frac{\mathrm{e}^{\mathrm{i}k_a r_j}}{4\pi k_a^2r_j^3}\left[(k_a^2r_j^2+\mathrm{i}k_ar_j-1)\mathbb{1}\right.\\\nonumber
    && \left.+(-k_a^2r_j^2-3\mathrm{i}k_ar_j+3)\frac{\mathbf{x}_j\otimes\mathbf{x}_j}{r_j^2}\right] \, ,
\end{eqnarray}
where $\mathbf{x}_{j=\alpha-\beta}\equiv\mathbf{r}_\alpha-\mathbf{r}_\beta$ are the separation vectors between the emitters, $r_j=|\mathbf{x}_j|$, and $k_a=\omega_a/c=2\pi/\lambda$.



\textit{Flat dispersion relation.} We first focus on the situation where all components of the Green's tensor are translationally invariant, e.g., on an infinite one-dimensional chain, or on a ring where the dipoles are pointing perpendicularly to the ring's surface. Here, a Fourier transform $    \Tilde{G}_0(k,k_a)=\sum_{j} \mathrm{e}^{-\mathrm{i}kx_j}\bar{G}_0(\mathbf{x}_j,k_a) $, with $x_j=jd$ and $k\in[0,2\pi/d]$ diagonalizes the Green's tensor. We will focus specifically on the Green's tensor's component $G^\delta_0(\mathbf{x}_j,k_a)$ relevant when the dipole moments form an angle $\delta$ with respect to the chain. 
The real and imaginary parts of this Fourier transformed component are proportional to the energies and decay rates of the eigenmodes in the system, $V^\delta_k$ and $\Gamma^\delta_k$, respectively.


Let us start by analyzing the collective decay rates $\Gamma^\delta_k$ for an infinite one-dimensional chain [Fig.~\ref{fig:fig2}(a)]. Independently of $\delta$, $\Gamma^\delta_k=0$ for all values $2\pi/d-k_a>k>k_a$, i.e., all states that lie between the so-called light lines, are completely subradiant \cite{Needham2019,Asenjo2017}. Outside of these lines, i.e. $k_a\geq k\geq0$ and $2\pi/d\geq k\geq 2\pi/d-k_a$, the states are typically superradiant, i.e. $\Gamma^\delta_k>\gamma=k_a^3|\mathbf{d}|^2/(3\pi\hbar\epsilon_0)$. Since we are interested in creating excitations that are long-lived, we will focus only on the creation of states that have a large support with the states within this subradiant region.

We study now the real part of the Green's tensor in Fourier space, which yields the spectrum or dispersion relation 
shown in Fig.~\ref{fig:fig2}(b). For wave packets which are localized in $k$-space, the derivative of $V_k^\delta$ with respect to $k$ gives us the group velocity $v_g^\delta$ with which the wave packet travels in real space. 
This group velocity is always zero at $k=k_0=\pi/d$ and $k=0$, as imposed by the continuity of the dispersion relation at the border of the first Brillouin zone. As explained in detail in the Supplemental Material \cite{SM}, in order to find an approximate flat dispersion relation for a large (and subradiant) section of the spectrum, typically hard to obtain in photonic systems \cite{Settle2007,Xu2007}, we calculate the values of $d/\lambda$ where $k=k_0$ becomes a saddle point. As displayed in Fig.~\ref{fig:fig2}(c), for all values of $\delta\in\left(\arccos{1/\sqrt{3}},\pi/2\right]$ we obtain a $d_\mathrm{f}(\delta)$ that leads to such a flat dispersion relations [illustrated in Fig.~\ref{fig:fig2}(b) for $\delta=\pi/2$ and the inset of Fig.~\ref{fig:fig2}(c) for $\delta=3\pi/8$]. In particular, we find $d_\mathrm{f} \equiv d_\mathrm{f}(\pi/2)=0.2414\lambda$ [see Fig.~\ref{fig:fig2}(c)], which has also been found to maximize the lifetime of the subradiant states \cite{Kornovan2019,Zhang2020flatband,Kornovan2021}. Note that similar results can be found for a ring lattice with $\delta=\pi/2$, a translationally symmetric problem also for a finite number $N$ of emitters. 
Here, subradiant states frozen in real space may be already created for very small system sizes (see Supplemental Material \cite{SM}). 

Our aim is now to show that the existence of a flat dispersion relation allows for high-fidelity and subradiant storage of a photon. To do so, we investigate the dynamics of a single-photon wave packet. In $k$-space, one can write such a state generally as $\ket{\Psi_0}=\sum_{k=0}^{2\pi/d} f(k,k_s,\sigma) \ket{k},$ where $f(k,k_s,\sigma)$ is a function centered around $k_s$ and width $\sigma$, with $2\pi/d-k_a>k_s>k_a$ and $\sigma\ll2|\pi/d-k_a|$ in order to ensure that the wave packet is subradiant. In particular, if we consider a Gaussian wave packet, i.e. $f(k,k_s,\sigma)=\mathrm{e}^{-\frac{(k-k_s)^2}{4\sigma^2}}/\sqrt{\sqrt{2\pi}\sigma}$, the wave packet in real space is also a Gaussian that reads 
\begin{equation}\label{eq:wavep}
    \ket{\Psi_0}=\sqrt{\frac{\sigma}{\sqrt{2\pi}}}\sum_{\alpha=1}^{N}\mathrm{e}^{-ik_sx_\alpha}\mathrm{e}^{-x_\alpha^2\sigma^2}\ket{e_\alpha} \, .
\end{equation}
Note that we have chosen such wave packet for illustration purposes. In general, if the wave packet is centered around $k_s$ with a small width in Fourier space, it will be subradiant and its time-evolution dispersionless. In the Supplemental Material \cite{SM}, the high-fidelity storage is demonstrated for a wave packet excited by a Gaussian laser beam.

\begin{figure}[t]
    \centering
    \includegraphics[]{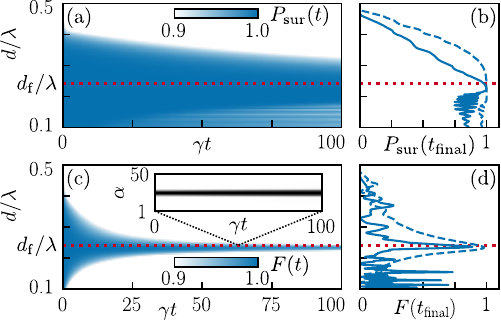}
    \caption{\textbf{Single-photon time evolution.} (a): Survival probability as a function of $d/\lambda$ and time $t$ of a photon stored as a wave packet (\ref{eq:wavep}) with $\sigma = 0.1 \pi / d$ and $\delta=\pi/2$ on a lattice with $N=50$ emitters. (b): $P_\mathrm{sur}(t_\mathrm{final})$ for $\gamma t_\mathrm{final}=100$ (dashed line) and $\gamma t_\mathrm{final}=500$ (solid line). (c-d): Same as (a-b) for the fidelity $F(t)$. Inset in (c): Time evolution of $\left<n_\alpha\right>$ at $d = d_\mathrm{f}$.}
    \label{fig:fig3}
\end{figure}

The time evolution of the initial state $\rho_0\equiv\rho(t=0)=\ketbra{\Psi_0}{\Psi_0}$ is obtained by solving the master equation (\ref{eq:meq}) on a finite one dimensional lattice with $N$ emitters in the single excitation regime \cite{Needham2019}. In order to evaluate the subradiant character of the storage, we calculate the survival probability of the photon, defined as
\begin{equation}
    P_\mathrm{sur}(t)= \sum_{\alpha=1}^N \operatorname{Tr} \left[\rho_t n_\alpha\right]=\sum_{\alpha=1}^N \expval{n_\alpha}_t \,,
\end{equation}
with $n_\alpha=\ketbra{e_\alpha}{e_\alpha}$ and $\rho_{t}\equiv\rho(t)$. For a single atom in free space, this survival probability is a decaying exponential, $P_\mathrm{sur}(t)=\mathrm{e}^{-\gamma t}$. We moreover evaluate the degree of dispersion of the initial wave packet by means of the fidelity \cite{Jozsa1994}, defined here as
\begin{equation}
    F(t)= \left[ \operatorname{Tr} \sqrt{\sqrt{\rho_0} \rho_t \sqrt{\rho_0}}\right]^2 \, .  
\end{equation}
In Fig.~\ref{fig:fig3} we show the time evolution of the survival probability and the fidelity for an initial wave packet with $k_s=k_0=\pi/d$ and $\sigma=0.1\pi/d$ created on a finite one-dimensional lattice of $N=50$ emitters and $\delta = \pi/2$, varying the ratio between the lattice spacing and the light's wavelength, $d/\lambda$. In order to avoid edge effects, the wave packet is created at the center of the lattice. One can observe in Fig.~\ref{fig:fig3}(a) that the survival probability reaches a maximum --$P_\mathrm{sur}(t_\mathrm{final})=0.9997$ for $t_\mathrm{final}=100/\gamma$--, ensuring a subradiant storage, provided that $d/\lambda$ is close to the value $d_\mathrm{f}/\lambda$ predicted above \cite{Kornovan2019}. While the range of $d/\lambda$ values for which $P_\mathrm{sur}(t_\mathrm{final})\approx1$ is quite broad at $t_\mathrm{final}=100/\gamma$, this distribution becomes even more peaked around $d_\mathrm{f}/\lambda$ for longer times [see Fig.~\ref{fig:fig3}(b)]. On the other hand, in Figs.~\ref{fig:fig3}(c) and (d), we can see that the fidelity is already maximum $F(t_\mathrm{final})=0.98$ around $d_\mathrm{f}/\lambda$ even at $t_\mathrm{final}=100/\gamma$. Hence, we have shown that a lattice constant $d=d_\mathrm{f}(\delta)$ allows for an optimal subradiant and dispersionless photon storage.

\begin{figure}[t]
    \centering
    \includegraphics[]{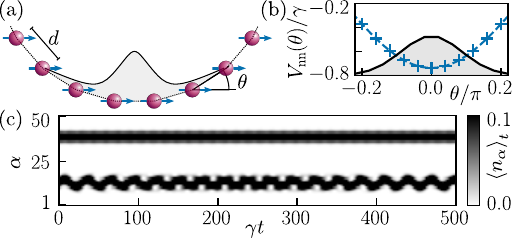}
    \caption{\textbf{Trapped states.} (a): In an emitter ring, the angle $\theta$ between the dipole moments and the separation vectors varies with the position of the emitter. (b): Nearest neighbour interactions $V_\mathrm{nn}$ for a ring of $N=50$ emitters and $d=0.234\lambda$ around $\theta=0$. (c): Time evolution of two initial wave packets, centered at $k_s = k_0$ and $k_s = k_0 (1 - 3 / 25)$ (above and below, respectively) and width $\sigma = 0.103 \pi/d$. Note that $d=0.234\lambda$ is not a necessary condition for the trapping, which can also be demonstrated for other values of $d$.}
    \label{fig:fig4}
\end{figure}

\textit{Trapped states.} We will now present a second storage mechanism of the single-photon state (\ref{eq:wavep}), which is not based on the presence of a flat dispersion relation. Given the overall form of the Green's tensor \eqref{eq:G0}, both coherent and incoherent interactions do not only depend on the reduced distance $r/\lambda$ between the emitters, but also on the angle $\theta$ between the two dipoles and the vector that separates them. Let us consider a ring where the dipole moments are contained in the plane of the ring [see Fig.~\ref{fig:fig4}(a)]. Here, the angle $\theta$ and hence the nearest neighbor interactions change as a function of the position on the ring. In particular, in the vicinity of $\theta=0$ these interactions can be approximated as $V_\mathrm{nn}(d,\theta)\approx -3\gamma/4 \left\{\left[A(d)-3B(d)\right]\theta^2+2B(d)\right\}$, with $A(d)=\cos{k_a d}/(k_a d)$ and $B(d)=\cos{k_a d}/(k_a^3 d^3)+\sin{k_a d}/(k_a^2 d^2)$, an approximately quadratic potential whose depth depends on the value of $d/\lambda$ [see Fig.~\ref{fig:fig4}(b)]. Making the simplifying assumption that only nearest neighbors interact, the Hamiltonian of the system can be approximated as a tight binding model
\begin{equation}\label{eq:TB}
    H\approx H_\mathrm{TB}=\sum_j V_\mathrm{nn}(d,\theta_j)\left(\ketbra{j}{j+1}+\mathrm{h.c.}\right) \, .
\end{equation}
We find that the Gaussian wave packet (\ref{eq:wavep}) with $k_s=k_0$ has a large overlap with one of the eigenstates of $H_\mathrm{TB}$. This overlap can be maximized choosing an optimal value of the width $\sigma$. Here, the state (\ref{eq:wavep}) becomes effectively an eigenstate of $H_\mathrm{TB}$, while being almost completely subradiant. Consequently, as we show in Fig.~\ref{fig:fig4}(c) (above), the wave packet remains trapped while keeping its shape orders of magnitude longer than the single emitter lifetime. Similarly, for a value of $k_s$ slightly deviating from $k_0$, the wave packet can be decomposed as a superposition of a few eigenstates of $H_\mathrm{TB}$, which can be identified in the dynamics by the dispersion and eventual revival of the wave packet [Fig.~\ref{fig:fig4}(c) below]. Note that here the dynamics is obtained by solving the (exact) master equation (\ref{eq:meq}), including all long-range interactions.

\begin{figure}
    \centering
    \includegraphics[width=\columnwidth]{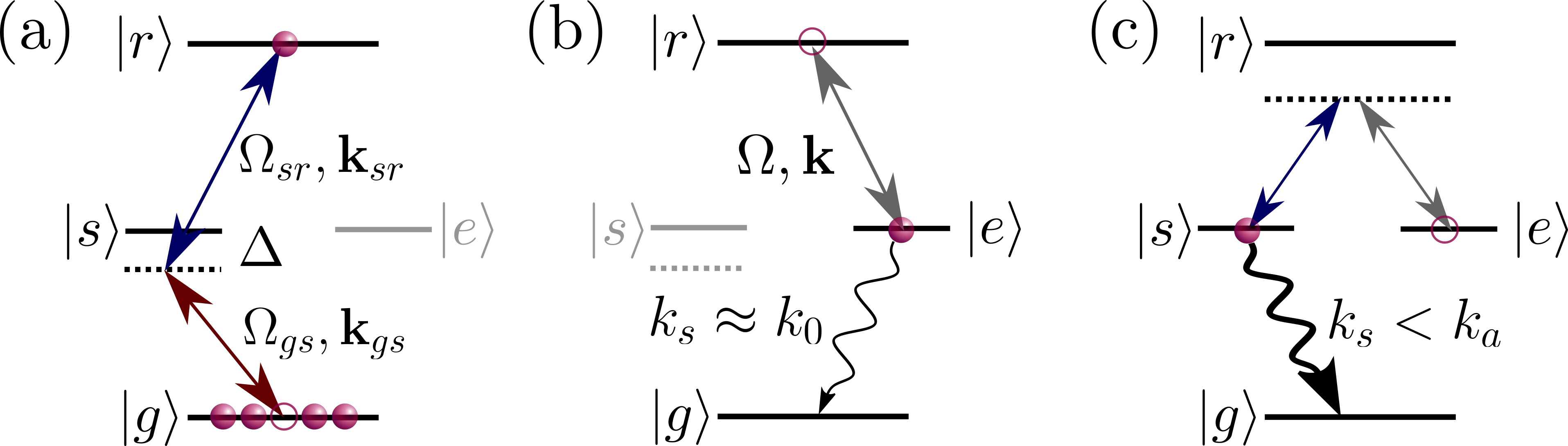}
    \caption{\textbf{Subradiant excitation creation and release.} Creation of the subradiant single photon state (\ref{eq:wavep}) with $k_s\approx k_0$ realized in two steps: (a) Laser excitation of a single-photon spin wave exploiting the strong interactions between atoms on a Rydberg state $\ket{r}$ and (b) mapping of this spin wave to the low-lying excited state $\ket{e}$. (c): The subradiant state (\ref{eq:wavep}) is Raman transferred into a superradiant one with $k_s<k_a$, outside the light lines.}
    \label{fig:fig5}
\end{figure}

\textit{Subradiant state preparation and release.} Finally, we propose a scheme for the laser excitation of the initial state (\ref{eq:wavep}). We face two challenges: to efficiently store exactly one photon avoiding the absorption of a second one, and to imprint a central momentum $k_s\approx k_0$, to ensure its subradiant character. To achieve this, we will consider two additional levels for each atom: a Rydberg state $\ket{r}$, with a high principal quantum number $n\gg1$, and an intermediate low-lying state $\ket{s}$ [see Fig.~\ref{fig:fig5}(a)]. Two laser fields drive the $\ket{g}\to\ket{s}$ and $\ket{s}\to\ket{r}$ transitions, with Rabi frequencies and momenta $\Omega_{gs}$, $\Omega_{sr}$ and $\mathbf{k}_{gs}$, $\mathbf{k}_{sr}$, respectively. The $\ket{s}$ state is far detuned ($\Delta\gg\Omega_{gs},\Omega_{sr}$), such that the ground state is coupled to the Rydberg state via a two-photon transition. 

Thanks to the strong long-ranged interactions between atoms in a Rydberg state \cite{Jaksch2000,Lukin2001,Urban2009}, inside an area with $N_b$ atoms determined by the so-called blockade radius $r_b$ (typically much larger than $d$), only one Rydberg excitation can exist. Hence, a $\pi$-pulse on this two-photon transition for a time $\tau_r=\pi/\Omega_\mathrm{eff}$, with $\Omega_\mathrm{eff}=\sqrt{N_b}\Omega_{gs}\Omega_{sr}/{2\Delta}$ will produce the state $\ket{\Psi_r}=\sum_{\alpha=1}^{N_b} \mathrm{e}^{-x_\alpha^2\sigma^2} \mathrm{e}^{-i (\mathbf{k}_{gs}+\mathbf{k}_{sr})\cdot\mathbf{x}_\alpha}\ket{r_\alpha}/\sqrt{N_b}$ which contains exactly one excitation (for the effects of using a more realistic Gaussian beam we refer the reader to the Supplemental Material \cite{SM}). The large lifetimes of Rydberg states (compared with the ones of typical low-lying electronic levels), allow us to assume that $\ket{\Psi_r}$ is stable. The second step is to map this state into (\ref{eq:wavep}), performed by another $\pi$-pulse with a laser that couples resonantly the Rydberg state to the excited state $\ket{e}$ with Rabi frequency $\Omega$ and momentum $\mathbf{k}$ [see Fig.~\ref{fig:fig5}(b)]. 
For the wave packet (\ref{eq:wavep}) to be subradiant, the momenta must satisfy that $\frac{2\pi}{d}-k_a>k_s=k_{gs}\cos{\theta_{gs}}+k_{sr}\cos{\theta_{sr}}-k\cos{\theta}>k_a$, where $\theta_{gs}$, $\theta_{sr}$ and $\theta$ are the angles formed by the interatomic vector $\mathbf{x}_j$ and the laser momenta. These parameters can be easily adjusted to make $k_s$ lie between the light lines. E.g., for Rb atoms with $\lambda=2\pi/k_{gs}=780$\,nm and $\lambda_r=2\pi/k_{sr}=2\pi/k=480$\,nm, $\theta=\pi/2$ and $\theta_{gs}=\theta_{sr}\approx2\pi/9$ one obtains $k_s\approx\pi/d_\mathrm{f}$.

Finally, the stored photon may be released by similarly transferring the central momentum outside the light lines. Here, a Raman transfer between the $\ket{e}$ and the $\ket{s}$ state via the Rydberg state [see Fig.~\ref{fig:fig5}(c)] produces a spin wave with central momentum $k_{gs}\cos{\theta_{gs}}<k_a$ that preserves the narrow width of the original subradiant state in momentum space, hence becoming a superradiant state, quickly lost via spontaneous decay within a narrow cone around the direction $\mathbf{k}_{gs}$ \cite{Masson2020,Petrosyan2021}.

\textit{Conclusions and outlook.} In this work, we have demonstrated through two different mechanisms the potential of one-dimensional sub-wavelength emitter chains for high-fidelity single-photon storage. Particularly appealing is the photon trapping exploiting the angular dependence of the dipole-dipole interactions, which provides extremely high fidelities and lifetimes (e.g., for the $N=50$ case explored in Fig.~\ref{fig:fig4}, $F(t_\mathrm{final})=0.998$ and $P_\mathrm{sur}(t_\mathrm{final})=0.999$ for $t_\mathrm{final}=500/\gamma$), while being quite robust against disorder in the emitter positions and specific shape of the laser beams involved in the state preparation and read-out (see Supplemental Material \cite{SM}\vphantom{\cite{Asenjo2017,Zhang2020flatband,svelto2010}}). As shown in \cite{Nikolopoulos2004,Christandl2004,Bose2007}, a finer-tuned choice of $V_\mathrm{nn}(d,\theta_j)$ gives rise to an approximate tight binding model (\ref{eq:TB}) that may allow for a fully non-dispersive transport of the wave packet across the lattice. Since $V_\mathrm{nn}(d,\theta_j)$ depends on the external geometry of the system, it will be interesting to investigate spatial arrangements of the emitters that not only optimize the fidelity and lifetime, but that also allow for the transport of the single-photon wave packet for long distances. Extending this scheme to more than one photon, such that, for example, two photons can be stored, transported, interact, and released, will also be explored.

The code and the data that support the findings of this Letter are available on Zenodo \cite{ZenodoData}.

\acknowledgments
\textit{Acknowledgements.} The research leading to these results has received funding from the Deutsche Forschungsgemeinschaft (DFG, German Research Foundation) under Projects No. 435696605, 449905436 and 452935230, as well as through the Research Unit FOR 5413/1, Grant No. 465199066.

\bibliography{references.bib}


\onecolumngrid
\clearpage

\subfile{supplemental_material}


\end{document}

%% file: supplemental_material.tex

\setcounter{equation}{0}
\setcounter{figure}{0}
\setcounter{table}{0}
\setcounter{page}{1}
\makeatletter
\renewcommand{\theequation}{S\arabic{equation}}
\renewcommand{\thefigure}{S\arabic{figure}}
\renewcommand{\bibnumfmt}[1]{[S#1]}
\renewcommand{\citenumfont}[1]{S#1}
\onecolumngrid
\setcounter{page}{1}

\begin{center}
{\Large SUPPLEMENTAL MATERIAL}
\end{center}
\begin{center}
\vspace{0.8cm}
{\Large Dispersionless subradiant photon storage in one-dimensional emitter chains}
\end{center}
\begin{center}
Marcel Cech\,\orcidlink{0000-0002-9381-6927},$^{1}$ Igor Lesanovsky\,\orcidlink{0000-0001-9660-9467},$^{1,2}$ and Beatriz Olmos\,\orcidlink{0000-0002-1140-2641}$^{1,2}$
\end{center}
\begin{center}
$^1${\em Institut f\"ur Theoretische Physik, Universit\"at T\"ubingen,}\\
{\em Auf der Morgenstelle 14, 72076 T\"ubingen, Germany}\\
$^2${\em School of Physics and Astronomy and Centre for the Mathematics\\ and Theoretical Physics of Quantum Non-Equilibrium Systems,\\ The University of Nottingham, Nottingham, NG7 2RD, United Kingdom}
\end{center}


\section{I. Calculation of $d_\mathrm{f}(\delta)$ for flat dispersion relation}
In this section, we detail the derivation of the lattice spacing $d_\mathrm{f}(\delta)$ to have an approximate flat dispersion relation at different orientation angle $\delta$ of the dipole moments with respect to the chain. We start by considering the dispersion relation
\begin{align}
    \label{eq:V_k_delta}
    \begin{split}
        V_k^\delta = \frac{3 \gamma}{4 k_a^3d^3} \operatorname{Re} \Bigl[& (1-3\cos^2\delta) \left( \operatorname{Li}_3(\mathrm{e}^{\mathrm{i}(k_a + k)d}) + \operatorname{Li}_3(\mathrm{e}^{\mathrm{i}(k_a - k)d}) - \mathrm{i}k_a d \operatorname{Li}_2(\mathrm{e}^{\mathrm{i}(k_a + k)d}) - \mathrm{i}k_a d \operatorname{Li}_2(\mathrm{e}^{\mathrm{i}(k_a - k)d}) \right)\\
        & + \sin^2\delta \left(k_a^2 d^2 \operatorname{Ln} (1-\mathrm{e}^{\mathrm{i}(k_a + k)d}) + k_a^2 d^2 \operatorname{Ln} (1-\mathrm{e}^{\mathrm{i}(k_a - k)d}) \right) \Bigr] \,
    \end{split}
\end{align}
for an emitter chain of lattice constant $d$ and dipoles oriented at an site-independent angle $\delta$ \cite{Asenjo2017}. In this expression, $\operatorname{Li}_n(x)$ denotes the polylogarithm of order $n$. We now calculate the second derivative of Eq.\,(\ref{eq:V_k_delta}) with respect to $k$, which essentially encodes the change of the group velocity $v_k^\delta$, at $k = k_0 = \pi / d$ (the end/beginning of the Brillouin zone) and find \cite{Zhang2020flatband}
\begin{align}
    \label{eq:change_in_v_g_delta}
    \frac{\partial}{\partial k} v_k^\delta \Bigr|_{k_0} = \frac{3\gamma d}{2 k_a} \left[ (1 - 3 \cos^2\delta) \left( \operatorname{Ln}\left(2 \cos \frac{k_a d}{2}\right) + \frac{k_a d}{2} \tan(\frac{k_a d}{2}) \right) - \sin^2\delta \left( \frac{k_a d}{2} \right)^2 \frac{1}{\cos^2(\frac{k_a d}{2})} \right] \, .
\end{align}
We now set this expression to zero to find the condition for an approximate flat dispersion relation. Despite the analytic form of the expression above, the equation is transcendental and hence we calculate the solutions numerically.

\begin{figure}[ht]
    \centering
    \includegraphics{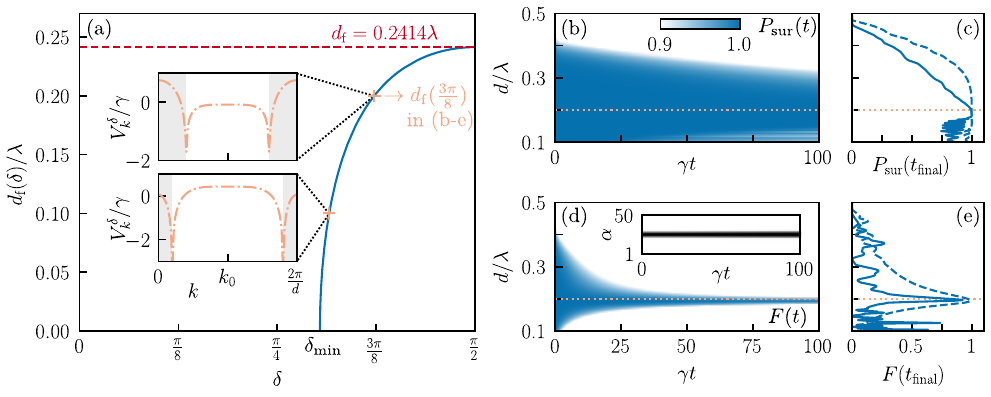}
    \caption{\textbf{Lattice spacings with flat dispersion relation.} {(a):} Parameter pairs $\{\delta, d_\mathrm{f}(\delta)\}$ for which Eq.\,(\ref{eq:change_in_v_g_delta}) vanishes. The lattice spacing $d_\mathrm{f} = 0.2414\lambda$ (red dashed line) providing a flat dispersion relation for perpendicular dipoles poses an upper bound for the lattice spacings with this feature. For two chosen lattice spacings $d_\mathrm{f}(\delta) = 0.1 \lambda, 0.2\lambda$, the insets show the dispersion relation with a flat section around $k_0 = \frac{\pi}{d}$ and well outside the radiative regime (gray shading). {(b-e):} Survival probability and fidelity, respectively, as shown in Fig.~3 in the main manuscript but for $\delta = 3\pi/8$ and $d_\mathrm{f}(\delta) = 0.2\lambda$ (orange dashed line).} 
    \label{fig:fig_s1}
\end{figure}

We visualize the results in Fig.~\ref{fig:fig_s1}(a). Here, we observe that there is a minimum angle $\delta_\mathrm{min}$ below which the equation does not have a solution. Taking the limit $d\to0$ in Eq.\,(\ref{eq:change_in_v_g_delta}), we find $\delta_\mathrm{min} = \arccos(1 / \sqrt{3})$. For $\delta\in\left(\delta_\mathrm{min},\pi/2\right]$, there exists a $d_\mathrm{f}(\delta)$ which leads to an approximate flat band, as one can see from the two insets in Fig.~\ref{fig:fig_s1}(a). The remarkably subradiant and dispersionless dynamics demonstrated in the main manuscript for $\delta=\pi/2$ persists for all these pairs of values, as illustrated in Fig.~\ref{fig:fig_s1}(b-e), where we investigate the survival probability and the fidelity for times up to $\gamma t_\mathrm{final}=100$ and $\gamma t_\mathrm{final}=500$, respectively.  

\section{II. Flat dispersion relation in finite systems}

\begin{figure}[ht]
    \centering
    \includegraphics{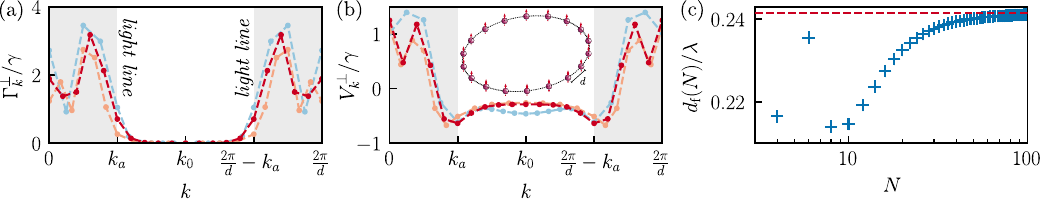}
    \caption{\textbf{Flat dispersion relation.} (a): Decay rates and (b): spectrum for a finite 1D ring lattice of $N=20$ emitters with lattice constant $d/\lambda=0.2$ (blue), $0.2414$ (red) and $0.26$ (orange), where the dipole moments are perpendicular to the surface of the ring. The same features as in Fig.~2 in the main manuscript (infinite 1D lattice) are found, i.e., subradiant modes between the light lines together with approximately flat spectrum around $k=k_0=\pi/d$. (c): Optimal lattice spacing $d_\mathrm{f}(N)$ for flat dispersion relation as a function of $N$. Already for small numbers of emitters $N$, it approaches the value for the infinite chain $d_\mathrm{f}=0.2414\lambda$ (red solid line).}
    \label{fig:fig_s2}
\end{figure}

In this section, we briefly discuss the applicability of arguments we made for the infinite chain to finite but periodic systems. We concentrate our study on $N$ emitters arranged on a ring lattice with dipole moments perpendicular to the ring plane ($\delta = \pi/2 \equiv \perp$).

Utilizing the same approach as for the infinite one-dimensional chain, we calculate the collective decay rate [Fig.~\ref{fig:fig_s2}(a)] and spectrum or dispersion relation [Fig.~\ref{fig:fig_s2}(b)] for a finite ring lattice of $N=20$ emitters. The direct comparison to Fig. 2(a-b) in the main manuscript shows a strong resemblance between the finite and infinite case. We find that subradiant states with finite lifetimes smaller than the single atom decay rate are found for momenta $2\pi/d-k_a>k>k_a$. Furthermore, the dispersion relation for these values of $k$ is approximately flat for values of $d$ close to the one found for the infinite lattice, $d_\mathrm{f}$. 

We quantitatively search for the optimal lattice spacing $d_\mathrm{f}(N)$ in the finite ring with $N$ emitters by numerically calculating the lattice spacing $d_\mathrm{f}(N)$ with a vanishing curvature of $V_k^\perp$ at $k=k_0$. In Fig.~\ref{fig:fig_s2}(c) we compare these values to $d_\mathrm{f}=0.2414\lambda$ found for the infinite chain. We observe that, as expected, already at small values of $N$, $d_\mathrm{f}(N)$ tends to the value of the infinite lattice.

\section{III. State preparation using a Gaussian beam}
For illustration purposes, we have primarily focused on the storage of a Gaussian wave packet [see Eq.\,(3)]. However, as we discussed in the main manuscript, for our storage scheme to work we only require that the wave packet stored has support mainly on the approximately flat area of the dispersion relation. Here, in particular, we demonstrate the storage of states that were prepared using a Gaussian laser beam.

\begin{figure}[t]
    \centering
    \includegraphics{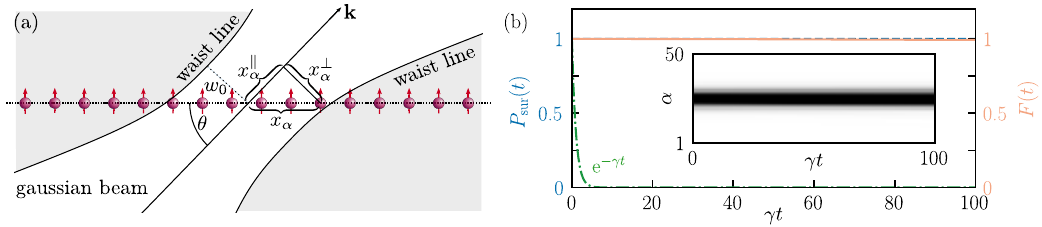}
    \caption{\textbf{State preparation using a Gaussian beam.} (a): Geometry of excitation with Gaussian beam. The Gaussian beam with wavevector $\mathbf{k}$ hits the chain of atom at an angle $\theta$ such that the cross-hair from its smallest waist $w_0$ and the beam center lies between the two central emitters. The waist lines illustrates where the electric field decreases to $1/\mathrm{e}$ of the axial value. $x_\alpha^\parallel$ ($x_\alpha^\perp$) represent the axial (perpendicular) decomposition of the atomic position $x_\alpha$. (b): Storage of wave packet prepared by Gaussian beam with waist $w_0 \approx 3.4 \lambda$. The waist is chosen such that the initial state has a width of approximately $\sigma = 0.06 \pi / d$. The one-dimensional chain of emitters is characterized by $d = d_\mathrm{f} = 0.2414\lambda$ and $N = 50$. We investigate the survival probability $P_\mathrm{sur}(t)$ and the fidelity $F(t)$ of the new, non-Gaussian initial state. $\mathrm{e}^{-\gamma t}$ corresponds to the single emitter survival probability. The inset underlines that we still obtain dispersionless subradiant storage.}
    \label{fig:fig_s3}
\end{figure}

The electric field at the emitters position $x_\alpha$ [see Fig.~\ref{fig:fig_s3}(a) for the decomposition into $x_\alpha^\parallel$ and $x_\alpha^\perp$] is given by 
\begin{align}
    \label{eq:gaussian_beam}
    E(x_\alpha) = E_0 \frac{w_0}{w(x_\alpha^\parallel)} \exp\left( \frac{-(x_\alpha^\perp)^2}{w(x_\alpha^\parallel)^2} \right) \exp \left( -i \left( k x_\alpha^\parallel  + k \frac{(x_\alpha^\perp)^2}{2R(x_\alpha^\parallel)} -\varphi(x_\alpha^\parallel) \right) \right)\, ,  
\end{align}
with the spot size $w(x_\alpha^\parallel) = w_0\sqrt{1 + \left( x_\alpha^\parallel / x_R \right)^2}$, the Rayleigh range $x_R = \pi w_0^2 n / \lambda$ (we set the refractive index $n=1$), the radius of curvature $R(x_\alpha^\parallel) = x_\alpha^\parallel \left( 1 + (x_R / x_\alpha^\parallel)^2 \right)$ and the Gouy phase $\varphi(x_\alpha^\parallel) = \arctan(x_\alpha^\parallel / x_R)$ \cite{svelto2010}. In this expression, the minimal waist $w_0$ and the wavevector $\mathbf{k}$ (such that $k = |\mathbf{k}| = 2\pi / \lambda$) are the two parameters that we control. 

In Fig.~\ref{fig:fig_s3}(b), we investigate the dispersionless subradiant storage of the new initial state for the chain of emitters at $d = d_\mathrm{f} = 0.2414 \lambda$ with perpendicular dipole moments. We observe that also this wave packet is stored over a much longer time-span than the single emitter's lifetime $1 / \gamma$. Comparing the survival probability $P_\mathrm{sur}(t)$ and fidelity $F(t)$ at $\gamma t_\mathrm{final} = 100$ for this initial state and a wave packet described by $k_s = k_0$ and $\sigma = 0.06 \pi / d$, we find again impressive storage capacities, with $P_\mathrm{sur}(t_\mathrm{final}) \approx 0.999$ and $F(t_\mathrm{final}) \approx 0.998$.

\section{IV. Disorder}
We give here a brief discussion on how robust the two mechanisms for long-lived and dispersionless storage are against positional disorder. Deviations of the emitter positions from a perfect lattice give rise to disorder in both the interaction and decay rates in Eq.~(1). We model the disorder by averaging over many realizations. In each realization we choose the position of the emitters randomly according to a three dimensional Gaussian centered on each lattice site with equal widths on all three directions, $\sigma_{d}$, which is in practice determined by the lattice depth. We investigate the influence of the disorder on the survival probability and the fidelity of both storage mechanisms.

\begin{figure}[ht]
    \centering
    \includegraphics{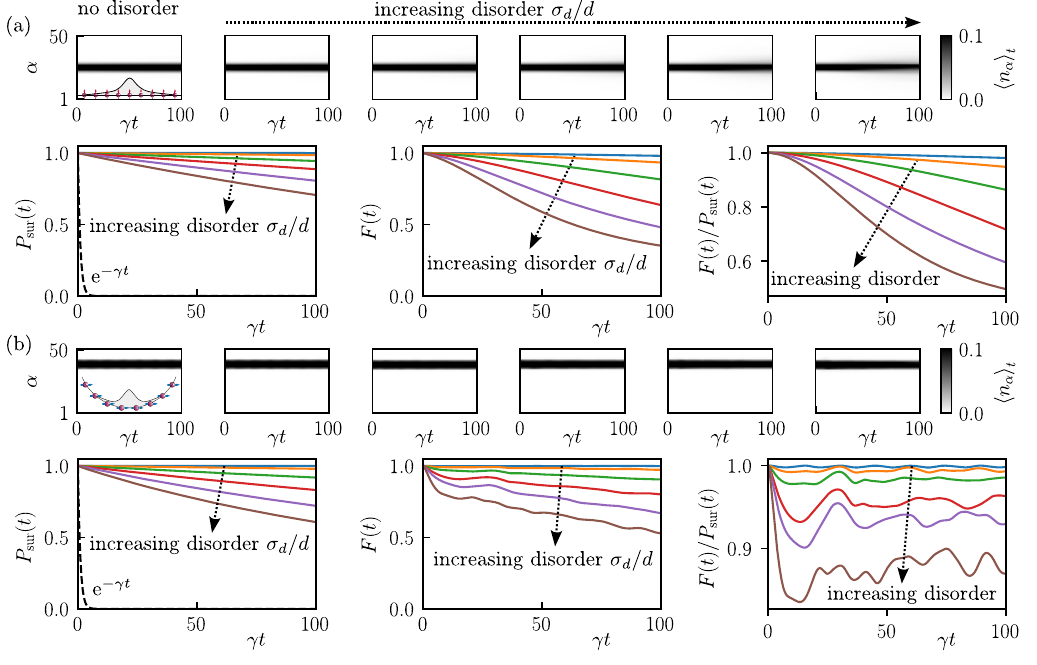}
    \caption{\textbf{Disorder.} Influence of increasing disorder on storage with (a): flat dispersion relation [see Fig.~3] and (b): trapped states [see Fig.~4]. Each panel compares the excitation probability $\left< n_\alpha \right>_t$ at site $\alpha$, the survival probability $P_\mathrm{sur}(t)$, the fidelity $F(t)$ and the ratio of latter $F(t) / P_\mathrm{sur}(t)$ two without disorder (leftmost panel, blue lines) to the averages over $100$ realizations of disorders characterized by the widths $\sigma_d = 0.01d,\,...,\,0.05d$. We set $d=0.234\lambda$ and investigate wave packets (3) with (a): $k_s=k_0$ and $\sigma=0.1\pi/d$ on a lattice with perpendicular dipoles of $N=50$ emitters and (b): $k_s=k_0$ and $\sigma=0.103\pi/d$ on a ring lattice of the same size with dipole moments aligned parallel to its surface.}
    \label{fig:fig_s4}
\end{figure}

Analyzing the results in Fig.~\ref{fig:fig_s4}, we find that both storage mechanisms exhibit a similar robustness to the disorder. With increasing disorder, the survival probability as well as the fidelity decreases in comparison to the regular lattice spacing. However, the storage is still notably enhanced over the single atom case. We also plot the ratio $F(t) / P_\mathrm{sur}(t)$, which represents the dispersion of the wave packet conditioned to the photon not having been emitted. This ratio remains particularly high for the trapped state with $k_s=k_0$ [see Fig.~\ref{fig:fig_s4} (b)], meaning that if the excitation is still in the system after a time $t$, the state in which the photon is stored will still be the wave packet (3).



